%% file: main.tex
\documentclass[lettersize,journal]{IEEEtran}

\usepackage{amsmath,amsfonts}
\usepackage{algorithmic}
\usepackage{algorithm}
\usepackage{array}
\usepackage{textcomp}
\usepackage{stfloats}
\usepackage{url}
\usepackage{verbatim}
\usepackage{graphicx}
\usepackage{cite}
\usepackage{hyperref}
\usepackage{xcolor}
\usepackage{tabularx}
\usepackage{longtable}
\usepackage{multirow}
\usepackage{enumitem}
\usepackage{threeparttable} 
\usepackage{booktabs}
\usepackage{wasysym} 
\usepackage{amssymb} 
\usepackage{pifont}
\usepackage{xurl}
\usepackage{subcaption}

\usepackage{tikz}

\begin{document}

\title{Automated Classification of Cybercrime Complaints using Transformer-based Language Models for Hinglish Texts}

\author{\IEEEauthorblockN{Nanda Rani\IEEEauthorrefmark{1}\textsuperscript{\ddag}, Divyanshu Singh\IEEEauthorrefmark{2}\textsuperscript{\ddag}, Bikash Saha\IEEEauthorrefmark{1}\textsuperscript{\ddag}, Sandeep Kumar Shukla\IEEEauthorrefmark{1}}

\IEEEauthorblockA{\IEEEauthorrefmark{1}Department of Computer Science and Engineering, Indian Institute of Technology Kanpur, India\\ Email: \{nandarani,bikash,sandeeps\}@cse.iitk.ac.in}

\IEEEauthorblockA{\IEEEauthorrefmark{2}C3ihub, IIT Kanpur, India\\
Email: divyanshus@c3ihub.iitk.ac.in}

\thanks{\textsuperscript{\ddag}Authors contributed equally to this work.}
}




\maketitle

\begin{abstract}

The rise in cybercrime and the complexity of multilingual and code-mixed complaints present significant challenges for law enforcement and cybersecurity agencies. These organizations need automated, scalable methods to identify crime types, enabling efficient processing and prioritization of large complaint volumes. Manual triaging is inefficient, and traditional machine learning methods fail to capture the semantic and contextual nuances of textual cybercrime complaints. Moreover, the lack of publicly available datasets and privacy concerns hinder the research to present robust solutions. To address these challenges, we propose a framework for automated cybercrime complaint classification. The framework leverages Hinglish-adapted transformers, such as HingBERT and HingRoBERTa, to handle code-mixed inputs effectively. We employ the real-world dataset provided by Indian Cybercrime Coordination Centre (I4C) during CyberGuard AI Hackathon 2024. We employ GenAI open source model-based data augmentation method to address class imbalance. We also employ privacy-aware preprocessing to ensure compliance with ethical standards while maintaining data integrity. Our solution achieves significant performance improvements, with HingRoBERTa attaining an accuracy of $74.41\%$ and an F1-score of $71.49\%$. We also develop ready-to-use tool by integrating Django REST backend with a modern frontend. The developed tool is scalable and ready for real-world deployment in platforms like the National Cyber Crime Reporting Portal. This work bridges critical gaps in cybercrime complaint management, offering a scalable, privacy-conscious, and adaptable solution for modern cybersecurity challenges.

\end{abstract}


\input{Section/Introduction.tex}

\input{Section/relatedwork}

\input{Section/methodology}

\input{Section/experimental_setup}

\input{Section/results}

\input{Section/discussion}

\input{Section/conclusion}





\end{document}

%% file: Section/Introduction.tex
\section{Introduction}
\label{sec:introduction}

The exponential growth of the Internet usage, digital transactions, and online social interactions has led to a corresponding surge in cybercrimes worldwide. Threats ranging from financial fraud~\cite{reurink2018financial,karpoff2021future}, ransomware~\cite{rani2022leveraging,rani2022survey}, and cyberterrorism~\cite{dawson2015brief}
to the proliferation of illicit content have become pervasive, placing immense pressure on law enforcement agencies and security analysts. Timely identification, classification, and prioritization of cybercrime complaints are now critical to effectively allocate investigative resources and prevent significant harm to individuals, organizations, and national infrastructures.

Recent studies show a surge in cybercrime in India and globally. In 2023, India reported 129 cybercrimes per lakh population~\cite{timesofindia}, and between January and April 2024, daily cybercrime complaints averaged $7,000$, a $113.7\%$ increase from $2021–2023$~\cite{businessstandard}. Globally, cybercrime costs are projected to rise by $15\%$ annually, reaching $\$10.5$ trillion by $2025$~\cite{norton}. Manual triaging of cybercrime complaints remains a complex and resource-intensive process, with human analysts struggling to cope with an ever-increasing volume of reports. Automated approaches to classify and prioritize these complaints have, therefore, attracted considerable research attention. 

The research in developing the automated model to understand the complaints has always struggled due to unavailability of public dataset of cyber crime complaints. Cybercrime complaints often contain sensitive and personal information, leading to stringent privacy concerns and limited data sharing. This scarcity of accessible datasets has hindered the development and evaluation of robust models, as researchers are unable to train and validate their approaches on diverse, real-world examples. Consequently, this gap restricts innovation and the ability to create universally effective solutions for automating and optimizing the handling of cybercrime complaints.

Additionally, early efforts relied on conventional machine learning (ML) techniques using Bag-of-Words or TF-IDF representations~\cite{he2009learning,kumari2018machine,nizamtext,andleeb2019identification}. However, these methods suffer from high dimensionality and increased complexity. They also fail to capture semantic nuances and context, especially in domains with specialized terminologies and complex scenario descriptions. For example, descriptions such as "distribution of CSAM (Child Sexual Abuse Material)" and "sharing explicit images of minors" represent the same criminal activities but use different terminologies. Bag-of-Words or TF-IDF treats these phrases as unrelated, failing to capture their semantic similarity and context. 

Moreover, in a country like India, cybercrime complaints are frequently composed in multilingual or code-mixed forms, posing an additional layer of complexity. In regions where Hindi and English coexist, Hinglish—a fluid blend of Hindi and English—commonly appears in user-generated content \cite{joshi2016measuring}. Such linguistic mixing challenges conventional NLP pipelines, as models trained solely on English corpora struggle to understand code-switching, lexical borrowing, and non-standard syntax.

Recent advancements in Natural Language Processing (NLP), particularly the advent of transformer-based language models such as BERT \cite{devlin2019bert} and RoBERTa \cite{liu2019roberta}, have revolutionized text classification tasks by providing contextualized representations~\cite{rani2023ttphunter,songailaite2023bert,ding2023airtag}. These models have demonstrated impressive performance on a wide range of NLP challenges. Yet, their effectiveness can wane in code-mixed settings where the training data does not reflect the language patterns of the target domain~\cite{nayak2022l3cube}. 

We address these gaps by presenting an automated crime complaints classification model based on transformer-based language model. To address the Hinglish-based linguistic mixing challenge by employing domain and language-adapted models, including HingBERT and HingRoBERTa, which are pre-trained on Hinglish corpora~\cite{nayak2022l3cube}. Specialized models better handle the linguistic complexities of code-mixed input, resulting in improved classification performance

We acknowledge the provision of the real-world crime complaint dataset by Indian Cybercrime Coordination Centre (I4C) during CyberGuard AI Hackathon 2024\footnote{\url{https://indiaai.gov.in/article/ai-for-citizen-safety-join-the-indiaai-cyberguard-ai-hackathon}}. The critical aspect of the provided dataset is the inherently imbalanced nature of the dataset. Commonly reported categories, such as financial fraud or social media crime, overshadow less frequently occurring but equally severe offenses like cyber trafficking or ransomware. Standard oversampling or undersampling techniques often fall short, risking distortion of feature distributions or under representation of minority classes~\cite{hassanat2022stop}. Therefore, we augment the dataset to increase the number of samples of minority class by generating new crime complaints samples. We use the advanced data augmentation framework powered by GenAI open source model, LLama3.1-7b, to generate contextually consistent synthetic samples~\cite{kobayashi2018contextual,karimi2021imbalanced,rani2024ttpxhunter}. The employed data augmentation method aids in enriching minority class instances and allowing models to generalize better across the entire crime categories.


Additionally, the ethical and privacy dimensions of cybercrime complaint classification cannot be overlooked. Complaints often contain sensitive, personally identifiable information (PII), necessitating stringent anonymization procedures to ensure compliance with data protection regulations and maintain user trust~\cite{lison2019anonymisation,rani2024comprehensive}. Careful preprocessing to replace such information with placeholders allows models to focus on the semantic and syntactic attributes of the complaints rather than memorizing individual entities. Therefore, we preprocess the data by employing regex pattern matching and entity replacement mechanism to replace the sensitive information with their corresponding entity type name as placeholder. 

\begin{figure}
    \centering
    \includegraphics[width=\linewidth, height=8.5cm]{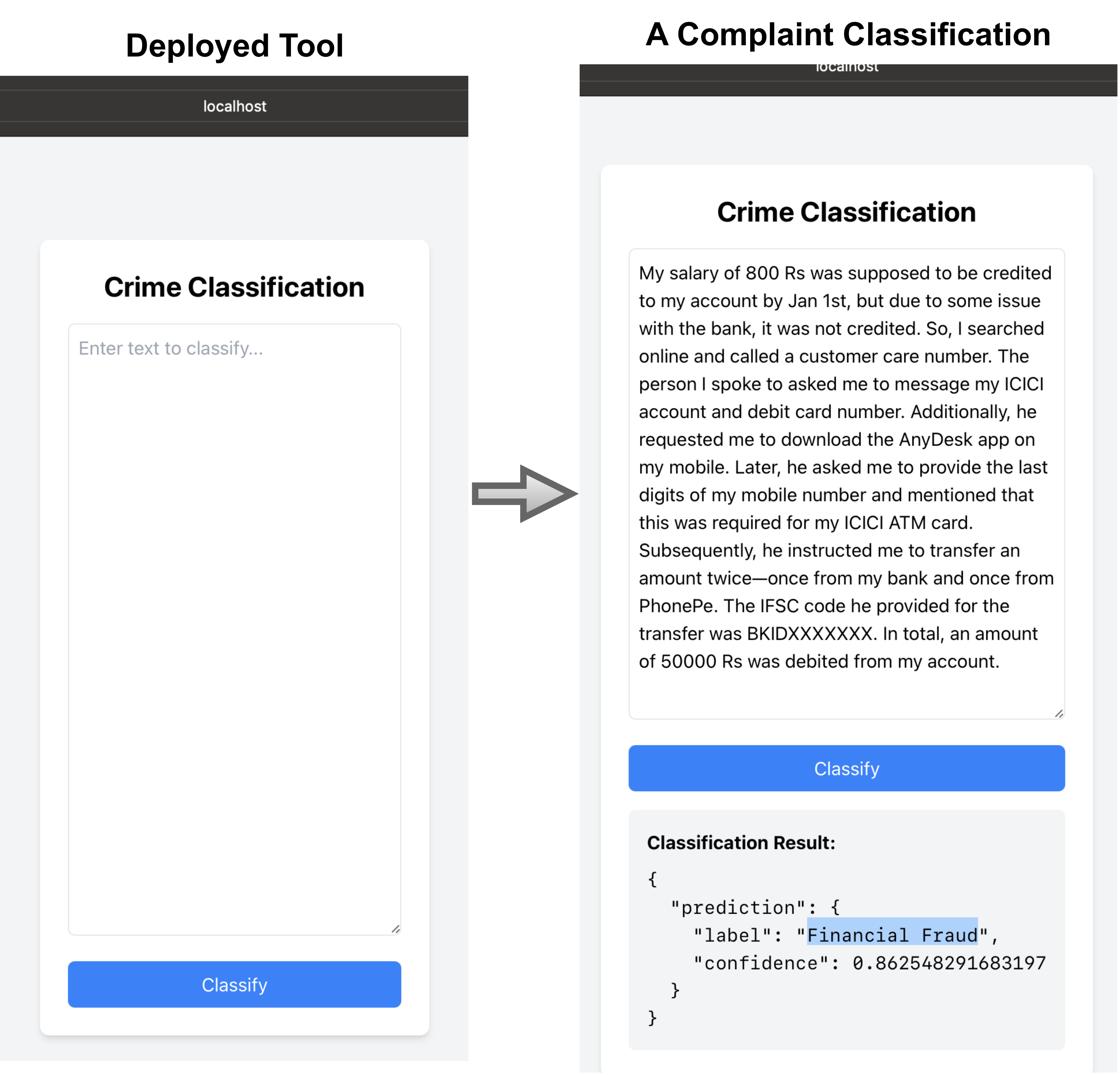}
    \caption{A Working Example of the Deployed Crime Classification Tool}
    \vspace{-1mm}
    \label{fig:TOOL_DEMO}
\end{figure}

Further, we evaluate the framework with traditional BERT and RoBERTa along with HingBERT and HingRoBERTa. The framework based on HingRoBERTa achieves significant performance improvements, by attaining the accuracy of $74.41\%$ and the F1-score of $71.49\%$. Furthermore, we integrate the solution with a Django REST backend and a modern JavaScript-based frontend, facilitating its deployment in real-world applications such as the National Cyber Crime Reporting Portal. This approach offers a scalable, adaptable, and privacy-conscious solution to the complexities of cybercrime complaint handling. The key contribution of this research are following:

\begin{itemize}
    \item We present a framework based on transformer-based language model to classify cyber crime complaints tuned specially to handle Hinglish texts.
    \item In the process of developing such solution, we also present the augmented dataset\footnote{We plan to release it publicly upon acceptance to support future research, addressing the current lack of publicly available crime complaint datasets for research.} of 25k cyber complaints distributed over $14$ crime type. The augmented dataset aligned with base dataset semantically and contextually.
    \item We present a ready-to-use classification tool\footnote{The tool will be available upon acceptance to assist future researchers in comparing state-of-the-art literature.}, based on our presented framework which can directly plug and play by the law enforcement agencies and other relevant organizations. A demo snapshot of our tool is shown in Fig~\ref{fig:TOOL_DEMO}.
    \item We also evaluate our framework with four different language model: BERT, RoBERTa, HingBERT, and HingRoBERTa, and perform comparison over various performance metrics to choose the best performing classifier.
\end{itemize}

\begin{table*}[]
    \centering
    \begin{tabular}{|p{1.8cm}|p{1.4cm}|p{1cm}|p{1.3cm}|p{1.3cm}|p{1.7cm}|p{1cm}|p{1.5cm}|p{1.5cm}|p{1cm}|}
    \hline
         \textbf{Paper} & \textbf{Data Source} & \textbf{Dataset Size} & \textbf{Data Augmentation} & \textbf{Feature} & \textbf{Model} & \textbf{Context Aware} & \textbf{Code-Mixed Aware} & \textbf{Privacy Compliance} & \textbf{Deployed Tool}\\
         \hline
         Sedik and Romadhony~\cite{sedik2023information} & News Articles & $1,963$ & \ding{55} & TF-IDF  & Support Vector Machine & \ding{55} & \ding{55} & \ding{55} & \ding{55}  \\
         \hline
         Islam et al.~\cite{islam2022machine} & News Articles & $5,318$ & \ding{55} & Count Vectorizer and TF-IDF & Random Forest & \ding{55} & \ding{55} & \ding{55} & \ding{55} \\
         \hline
         Rahma and Romadhony~\cite{rahma2021rule} & News Article & $40$ & \ding{55} & Ontology features & Rule-based & \ding{55} & \ding{55} & \ding{55} & \ding{55} \\
         \hline
         Prabhu et al.~\cite{prabhu2023cyber} & Cyber Complaints & $-$ & \ding{55} & Keyword Extraction & ERNIE classifier & \ding{55} & \ding{55} & \ding{55} & \ding{51} \\
         \hline
         Pongpaichet et al.~\cite{pongpaichet2024camelon} & News Articles & $8,567$ & \ding{55} & RoBERTa Embeddings & XLM-RoBERTa & \ding{51} & \ding{55} & \ding{55} & \ding{55} \\
         \hline
         \textbf{Our Method} & \textbf{Real-world crime complaints} & $\textbf{1,09,294}$ & \ding{51} & \textbf{Contextual Embeddings} & \textbf{HingRoBERTa} & \ding{51} & \ding{51} & \ding{51} & \ding{51} \\
         \hline
    \end{tabular}
    \caption{Comparison of Our Method with Current Literature}
    \label{tab:RelatedWorkComparison}
\end{table*}

\noindent The remainder of this paper is organized as follows: Section~\ref{sec:related_work} reviews related literature on cybercrime classification, code-mixed NLP, and data augmentation strategies. Section~\ref{sec:methodology} details our methodology, including preprocessing, model architectures, and augmentation methods. Section~\ref{sec:experiments} describes the experimental setup, while Section~\ref{sec:results} presents the results, evaluation and key discussions.
We conclude by outlining future directions in Section~\ref{sec:conclusion}.




%% file: Section/relatedwork.tex
\section{Related Work}
\label{sec:related_work}

The increasing prevalence of cybercrime has prompted extensive research into automated crime type classification. Existing studies have explored various methods, and textual languages, but significant gaps remain, particularly in handling multilingual and code-mixed inputs. In this section, we discuss recent advancements and their limitations, setting the stage for the contributions of our work.

Sedik and Romadhony~\cite{sedik2023information} proposed a system for crime information extraction from Indonesian crime news articles using a combination of Named Entity Recognition (NER) and machine learning models such as Support Vector Machines (SVM). While their work effectively categorizes crime types using TF-IDF features, it is limited by the reliance on handcrafted features and predefined crime categories. Additionally, their approach does not address multilingual or code-mixed contexts, which are prevalent in many regions, including India.

Islam et al.~\cite{islam2022machine} focused on classifying Bangla crime news using supervised machine learning models. They use traditional feature extraction methods like CountVectorizer and TfidfVectorizer to prepare the feature vector. These methods struggle with capturing semantic and contextual nuances in textual data, particularly in scenarios involving complex linguistic structures like code-mixing. Furthermore, their reliance on Bangla-specific tools limits generalizability to other languages or mixed-language scenarios.

Rahma and Romadhony~\cite{rahma2021rule} employed a rule-based system for crime information extraction, combining dependency parsing and ontology-driven classification. Although this method achieves reasonable performance in extracting crime-related entities, it encounters significant challenges in adapting rules for language-specific complexities. This rigid rule-based framework is unsuitable for dynamic, code-mixed complaints and does not leverage recent advancements in deep learning for enhanced contextual understanding.

Prabhu et al.~\cite{prabhu2023cyber} introduced a Cyber Complaint Automation System using the RAKE algorithm for keyword extraction and ERNIE for complaint classification. Their approach demonstrated the potential of integrating external knowledge for improved performance. However, it primarily focuses on predefined classes like fraud or hacking without addressing the inherent class imbalance in cybercrime datasets. Additionally, the lack of a multilingual or code-mixed capability reduces the system's applicability in diverse linguistic settings like Hinglish.

Pongpaichet et al.~\cite{pongpaichet2024camelon} introduced CAMELON, a methodology for classifying crime types based on online news articles annotated across seven distinct crime categories. Four deep learning models, including BiLSTM with Thai2Vec and transformer-based models like WangchanBERTa, MBERT, and XLM-RoBERTa, were fine-tuned for crime classification using binary cross-entropy loss and optimized training parameters. Out of all implemented models, XLM-RoBERTa achieves promising results. The method focus on news articles rather than complaints limits its real-world applicability to law enforcement systems. Moreover, the lack of integration with privacy-preserving mechanisms leaves concerns about ethical compliance and data security unaddressed.

Despite these advancements, existing approaches face several limitations. Most studies are rely on word-level feature extraction methods like bag-of-words or TF-IDF, traditional machine learning models, or rule-based frameworks, which fail to capture the semantic and contextual intricacies of code-mixed complaints. Moreover, the lack of robust strategies to address class imbalance and ensure privacy compliance further restricts their real-world applicability.

To address these gaps, we propose a framework for automated cybercrime complaint classification. It leverages Hinglish-adapted transformers like HingBERT and HingRoBERTa to handle code-mixed inputs. A Generative AI-based data augmentation method addresses class imbalance, while privacy-aware preprocessing ensures ethical data handling. The framework achieves significant performance improvements and a tool based on this framework is designed for adoption by law enforcement agencies. The tool can be integrated with scalable tools to support platforms like the National Cyber Crime Reporting Portal. This work bridges key gaps, offering a practical and adaptable solution for modern cybersecurity challenges. A comparison between our method and current literature is shown in Table~\ref{tab:RelatedWorkComparison}.

%% file: Section/methodology.tex
\section{Methodology}
\label{sec:methodology}

\begin{figure*}
    \centering
    \includegraphics[width=\linewidth]{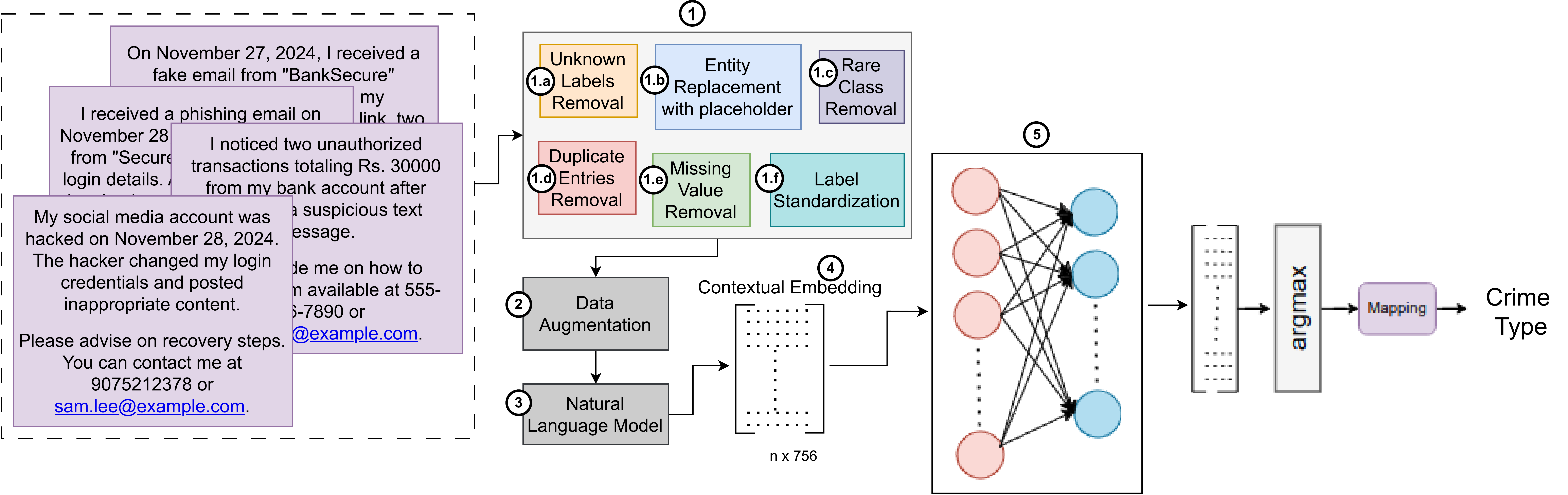}
    \caption{Architecture of Presented Crime Classification Method}
    \label{fig:methodology_arch}
\end{figure*}

Our overarching goal is to design a classification framework that can accurately and ethically process cybercrime complaints, many of which are written in code-mixed Hinglish and exhibit severe class imbalance. To achieve this, we integrate a pipeline of data preprocessing, language model adaptation, GenAI-based data augmentation, and fine-tuning strategies that collectively enhance the robustness, fairness, and applicability of the system. Figure~\ref{fig:methodology_arch} provides a high-level overview of our methodology.


\subsection{Problem Formulation and Requirements}
Given a dataset $\mathcal{D} = \{(x_i, y_i)\}_{i=1}^{N}$ of $N$ cybercrime complaints, where each complaint $x_i$ is a Hinglish text snippet and $y_i \in \{c_1, c_2, \dots, c_K\}$ is one of $K$ predefined classes, such as financial fraud and social media crime, we present a classification function $f: x \mapsto y$ that maximizes accuracy and F1-score across all categories. The presented solution also addresses practical constraints, including privacy awareness and seamless deployment.

\subsection{Dataset Preparation and Splits}
We obtain the real-world dataset from the Indian Cybercrime Coordination Center (I4C) during the CyberGuard AI Hackathon 2024. The dataset encompasses the total number $85,875$\footnote{After removing duplicates and other relevant preprocessing methods explained in Section~\ref{subsec:data_preprocess}} of crime complaints distributed in $14$ crime categories, as shown in Table~\ref{tab:dataset_distribution}. The dataset contains two data sets: training and testing dataset. We divide $20\%$ of the training data as validation set. The data set obtained is highly imbalanced, as shown in Table~\ref{tab:dataset_distribution} and contains complaints in the Hinglish. To handle imbalance data, we implement augmentation method on training dataset which we discuss in Section~\ref{subsec:data_augment}.

\subsection{Data Preprocessing}
\label{subsec:data_preprocess}
Effective data preprocessing is foundational for producing high-quality inputs that models can leverage. We begin by removing or adjusting problematic samples discovered through exploratory data analysis (EDA):

\begin{enumerate}
    \item \textbf{Unknown and Rare Classes:} Classes absent in the training set, such as \emph{Crime Against Women \& Children} found only in the test set, are excluded from the evaluation (step {\Large \textcircled{\small 1.a}} in Fig~\ref{fig:methodology_arch}). Classes with fewer than 2 samples in the training set, such as \emph{Report Unlawful Content}, are removed to ensure sufficient data for splitting into training and validation sets (step {\Large \textcircled{\small 1.c}} in Fig~\ref{fig:methodology_arch}).
    
    \item \textbf{Handling Missing and Duplicate Data:} Samples with critical fields, such as complaint description, containing NaN values or duplicates are removed to maintain dataset integrity and avoid skewed results (step {\Large \textcircled{\small 1.d}} and {\Large \textcircled{\small 1.e}} in Fig~\ref{fig:methodology_arch}).

    \item \textbf{Entity Anonymization:} Cybercrime complaints often contain sensitive and personally identifiable information (PII), such as names, emails, and addresses. We replace these entities with placeholder tokens (step {\Large \textcircled{\small 1.b}} in Fig~\ref{fig:methodology_arch}), such as \texttt{<Name>}, \texttt{<Email>}, and \texttt{<Location>}, to protect user privacy and ensure ethical compliance, enabling the model to learn generalizable patterns instead of memorizing unique entities. We use regex patterns to identify entities like phone numbers and emails and the Python library \emph{spacy} to detect entities such as names, addresses, and organization names. The list of entities and corresponding placeholder is shown in Table~\ref{tab:entity_place}.

    \begin{table}[!h]
    \centering
    \begin{tabular}{|c|c|c|}
    \hline
         \textbf{Entity Type} & \textbf{Utility} & \textbf{Placeholder}  \\
         \hline
         Name & Spacy & \texttt{<PERSON>} \\
         \hline
         Mobile Number & Regex & \texttt{<PHONE>} \\
         \hline
         Mail Address & Regex & \texttt{<EMAIL>} \\
         \hline
         Addresses & Spacy & \texttt{<ADDRESS>} \\
         \hline
         XYZ.com & Regex & \texttt{<WEBSITE>} \\
         \hline
         Monetary Value & Spacy & \texttt{<MONEY>}\\
         \hline
        \end{tabular}
        \caption{Entity Placeholder and used Utilities}
        \label{tab:entity_place}
    \end{table}

    \item \textbf{Stopword Removal and Lemmatization:} Common stopwords are removed to reduce noise, and words are lemmatized\footnote{A process of reducing a word to its base or dictionary form (lemma) while considering its context and part of speech.} to their base forms for linguistic consistency. This step encourages the model to focus on semantically significant terms.

    \item \textbf{Label Standardization:} The chosen dataset labels are lengthy, ambiguous, or overlapping. We standardize them (step {\Large \textcircled{\small 1.f}} in Fig~\ref{fig:methodology_arch}) to concise and clear labels to streamline learning and ensure distinct class boundaries. The standardization labels and their mapping is shown in Table~\ref{tab:label_standardization}.
\end{enumerate}

\begin{table*}[!ht]
\centering
\caption{Label standardization dictionary}
\begin{tabular}{|c|c|}
\hline
\textbf{Original Category} & \textbf{Simplified Category} \\ \hline
Any Other Cyber Crime & Other Cyber Crime \\ \hline
Child Pornography CPChild Sexual Abuse Material CSAM & Child Abuse Material \\ \hline
Cryptocurrency Crime & Cryptocurrency Crime \\ \hline
Cyber Attack/ Dependent Crimes & Cyber Attack/Dependent Crimes \\ \hline
Cyber Terrorism & Cyber Terrorism \\ \hline
Hacking Damage to computer computer system etc & Hacking/Damage \\ \hline
Online Cyber Trafficking & Cyber Trafficking \\ \hline
Online Financial Fraud & Financial Fraud \\ \hline
Online Gambling Betting & Gambling/Betting \\ \hline
Online and Social Media Related Crime & Social Media Crime \\ \hline
Ransomware & Ransomware \\ \hline
RapeGang Rape RGRSexually Abusive Content & Rape or Sexual Abuse Content \\ \hline
Sexually Explicit Act & Sexually Explicit Content \\ \hline
Sexually Obscene material & Sexually Obscene Content \\ \hline
\end{tabular}
\label{tab:label_standardization}
\end{table*}

These preprocessing steps yield a cleaner dataset that is more amenable to accurate classification and privacy-aware model training.

\subsection{Data Augmentation Method}
\label{subsec:data_augment}
\begin{figure*}[!h]
    \centering
    \includegraphics[width=13.5cm, height=6cm]{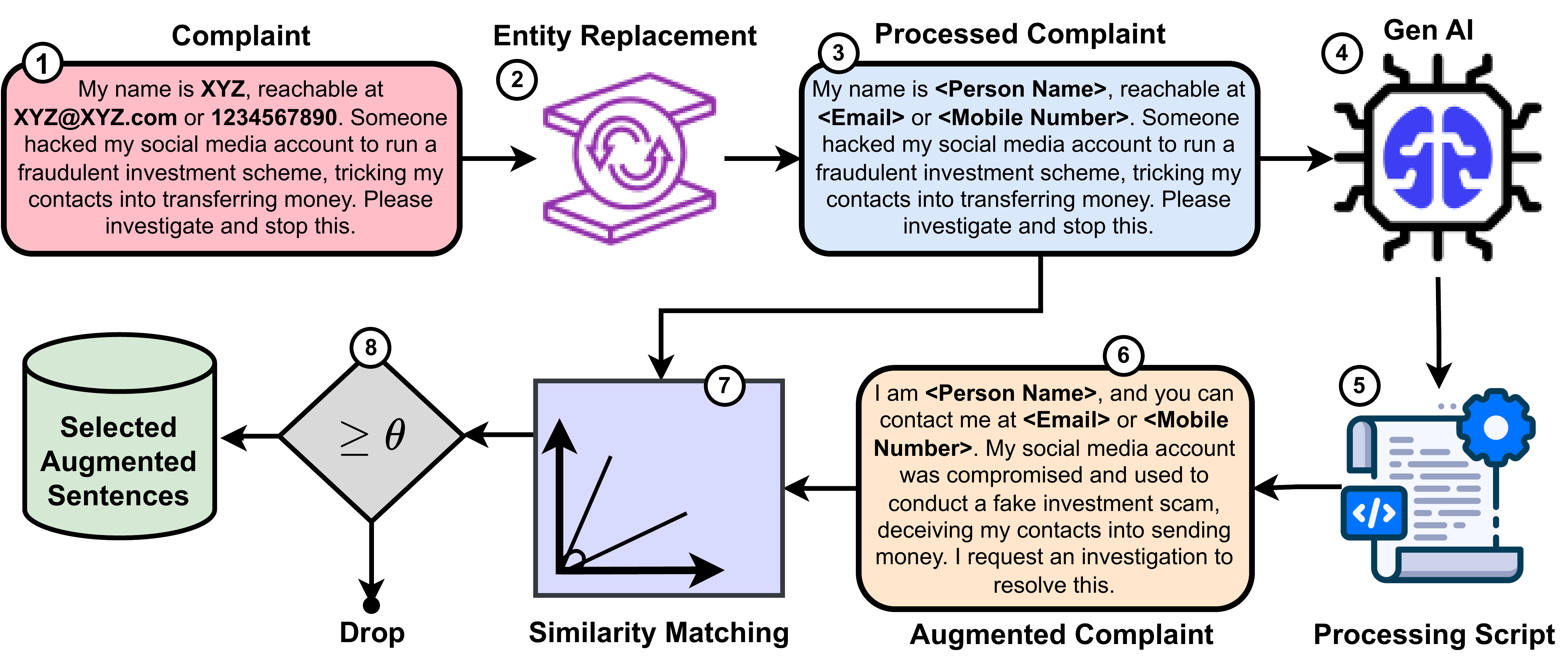}
    \caption{Data Augmentation Method}
    \label{fig:data_aug_arch}
\end{figure*}

One of the core challenges in our dataset is severe class imbalance, where a few dominant categories overshadow many minority classes. For example, $52,496$ out of $85,875$ samples belongs to Financial fraud. Conventional re-balancing techniques, such as naive oversampling or undersampling, often distort the data distribution or introduce artificial biases. Instead, we adopt a GenAI-based augmentation framework (step {\Large \textcircled{\normalsize 2}} in Fig~\ref{fig:methodology_arch}) inspired by our recent work on contextual augmentation \cite{kobayashi2018contextual,karimi2021imbalanced,rani2024ttpxhunter}. The architecture of augmentation method is shown in Fig.~\ref{fig:data_aug_arch} .

The process of data augmentation begins with preprocessing the dataset. Sensitive information, such as names, email addresses, and phone numbers, is replaced with placeholders like \texttt{<Person Name>, <Phone Number>} and other entities (step {\Large \textcircled{\normalsize 2}} in Fig~\ref{fig:data_aug_arch}) to ensure privacy and standardization. Once the data is prepared (step {\Large \textcircled{\normalsize 3}} in Fig~\ref{fig:data_aug_arch}), the Gen AI open-source model, LLaMA 3.1-7b, is used to generate augmented sentences (step {\Large \textcircled{\normalsize 4}} in Fig~\ref{fig:data_aug_arch}). The model creates paraphrased versions of the original sentences (step {\Large \textcircled{\normalsize 6}} in Fig~\ref{fig:data_aug_arch}) while maintaining their semantic and contextual meaning. Pre-processing steps such as removing excess spaces and new lines, segregating different examples as different augmented samples are performed (step {\Large \textcircled{\normalsize 5}} in Fig~\ref{fig:data_aug_arch}) to refine output sentences.

After generating the augmented data, a similarity evaluation is performed (step {\Large \textcircled{\normalsize 7}} in Fig~\ref{fig:data_aug_arch}). Each augmented sentence is compared to the original using semantic similarity measures to ensure relevance and contextual accuracy. To perform this evaluation, we use the BERTF1-score metric, which converts both sentences into contextual embeddings using a BERT layer and calculates the cosine similarity between the embeddings. Sentences that deviate significantly from the intended meaning or introduce ambiguity are discarded based on $\theta$ threshold (step {\Large \textcircled{\normalsize 8}} in Fig~\ref{fig:data_aug_arch}). For this experiment, we set a high threshold of $97\%$ to ensure that only highly similar and relevant sentences are filtered and included in the augmented dataset. The selected augmented sentences are then added to the dataset, enhancing its diversity and robustness. This expanded dataset is validated to ensure its effectiveness in improving the performance of downstream tasks.



By following this augmentation process, we increase the dataset size from $85,875$ to $1,09,294$ data samples. The distribution of the base dataset and augmented dataset is shown in Table~\ref{tab:dataset_distribution}. The augmented dataset gains a more balanced class distribution, which allow the model to develop a more equitable understanding of all crime categories.

\begin{table}[!ht]
\centering
\caption{Base and Augmented Dataset Distribution}
\begin{tabular}{|l|c|c|}
\hline
\textbf{Category}                     & \textbf{Base Count} & \textbf{Augmented Count} \\ \hline
Financial Fraud                       & 52496               & 52517                   \\ \hline
Social Media Crime                    & 12076               & 12086                   \\ \hline
Other Cyber Crime                     & 10811               & 10826                   \\ \hline
Cyber Attack/Dependent Crimes         & 3608                & 3625                    \\ \hline
Sexually Obscene Content              & 1764                & 6570                    \\ \hline
Hacking/Damage                        & 1709                & 8091                    \\ \hline
Sexually Explicit Content             & 1489                & 6118                    \\ \hline
Cryptocurrency Crime                  & 473                 & 2579                    \\ \hline
Gambling/Betting                      & 444                 & 2640                    \\ \hline
Child Abuse Material                  & 357                 & 1672                    \\ \hline
Rape or Sexual Abuse Content          & 248                 & 258                     \\ \hline
Cyber Trafficking                     & 183                 & 889                     \\ \hline
Cyber Terrorism                       & 161                 & 1150                    \\ \hline
Ransomware                            & 56                  & 273                     \\ \hline
\end{tabular}
\label{tab:dataset_distribution}
\end{table}

\subsection{Model Selection: Transformer-based Language Models}
Modern transformer-based language models are well-suited for tasks involving complex linguistic patterns and contextual dependencies. Traditional TF-IDF or Bag-of-Words representations fail to capture semantic relationships, especially in code-mixed text. We fine-tune four transformer-based models to capture semantic and contextual meaning for crime classification:

\begin{itemize}
    \item \textbf{BERT} \cite{devlin2019bert}: A foundational bidirectional transformer model widely used for various NLP tasks.
    \item \textbf{RoBERTa} \cite{liu2019roberta}: An enhanced version of BERT that focuses exclusively on the masked language modeling objective, often yielding improved results in classification tasks.
    \item \textbf{HingBERT} \cite{nayak2022l3cube}: A Hinglish-adapted BERT variant, pre-trained on code-mixed corpora. Its exposure to code-switching and multilingual vocabulary makes it more adept at handling Hinglish text.
    \item \textbf{HingRoBERTa}: A Hinglish-specific variant of RoBERTa, combining the advantages of robust optimization with Hinglish adaptation.
\end{itemize}

These models generate contextualized token embeddings, from which a sentence-level representation is extracted. Further, this representation is given to a fully connected linear classifier to predict the crime class label.

\subsection{Fine-tuning and Optimization}
We initialize the model with pre-trained weights and fine-tune it on the preprocessed, augmented dataset. Key hyperparameters include the learning rate, tuned through grid search within the range of $1e{-5}$ to $3e{-5}$, and batch size as 8 to 32 and sequence length as 128 or 256 tokens to accommodate most complaints without truncation. We use the AdamW optimizer for stable optimization, along with early stopping to prevent overfitting by halting training if validation performance stagnates for five epochs. A validation set aids in hyperparameter tuning and model checkpointing. While traditional ML models rely on k-fold cross-validation for robustness, transformer models utilize a stable validation split to balance computational efficiency with performance. We also use early stopping to prevent overfitting, halting training if validation accuracy or F1-score doesn’t improve for five epochs. This ensures balanced model complexity and reduces computational overhead.


\subsection{Deployment Architecture}
Beyond achieving good performance in a controlled experiment, we prioritize practical deployment. The final model is integrated into a Django REST-based backend that exposes RESTful APIs for classification requests. The frontend, built with modern JavaScript frameworks and Tailwind CSS, offers a user-friendly interface for operators to submit complaints and review automatically generated labels.

This production-ready architecture ensures scalability, monitoring, and easy integration with the National Cyber Crime Reporting Portal. The modular design allows for periodic model updates, retraining on new data, and the integration of more advanced models or additional categories as cybercrime patterns evolve.

\subsection{Summary of the Methodology}
In summary, our methodology strategically combines preprocessing, GenAI-based data augmentation, Hinglish-adapted transformer models, and careful optimization to build a robust and privacy-conscious classification system. By addressing code-mixing, class imbalance, and deployment concerns holistically, our approach delivers a practical, ethically responsible, and high-performing solution that can guide law enforcement efforts in navigating the complexities of modern cybercrime.

%% file: Section/experimental_setup.tex
\section{Experiments and Experimental Setup}
\label{sec:experiments}

In this section, we present the detailed configuration of our experiments, including the evaluation metrics, baseline setup, computational environment, and deployment configuration. Our objective is to ensure that all experiments are replicable, transparent, and sufficiently rigorous to draw meaningful conclusions about the proposed approach’s capabilities.

\subsection{Evaluation Metrics}
Given the complexity and imbalance inherent in the data, we consider multiple metrics:
\begin{itemize}
    \item \textbf{Accuracy:} Offers a broad performance overview but may be misleading in heavily imbalanced scenarios.
    \item \textbf{Precision, Recall, and F1-Score:} More informative measures for imbalanced classification. The F1-score, in particular, is critical because it harmonizes precision and recall, ensuring that good performance cannot be achieved by ignoring minority classes.
\end{itemize}

By employing a comprehensive metric set, we thoroughly examine each model’s strengths and weaknesses, ensuring that reported improvements truly reflect meaningful advances in handling code-mixed cybercrime complaints.

\subsection{Baseline Models and Comparisons}
Establishing robust baselines is essential for contextualizing performance gains. We compare our approach against:
\begin{itemize}
    \item \textbf{Traditional Machine Learning Pipelines:} We convert the text into TF-IDF features, apply dimensionality reduction via Singular Value Decomposition (SVD) to manage sparsity, and train models such as XGBoost, Random Forest, AdaBoost, and k-Nearest Neighbors. These pipelines serve as a reference point, showing how context-agnostic representations fare against our proposed methodologies.
    
    \item \textbf{Generic Transformer Models:} BERT \cite{devlin2019bert} and RoBERTa \cite{liu2019roberta} provide strong baselines from the transformer family. Their performance reveals whether context-aware embeddings alone (without Hinglish adaptation) suffice for the classification task.
    
    \item \textbf{Hinglish-Adapted Transformers:} HingBERT and HingRoBERTa \cite{nayak2022l3cube} are tested to confirm whether pre-training on code-mixed data confers distinct advantages. Improved performance from these models would validate the notion that language-adaptation is critical for handling Hinglish inputs effectively.
\end{itemize}

By comparing baselines with specialized models, we assess the impact of code-mixed language modeling and GenAI-based augmentation on classification performance.



\subsection{Hardware and Software Environment}
All experiments were performed on high-performance GPU servers equipped with NVIDIA Tesla or V100 accelerators. Running transformer fine-tuning with batch sizes of up to 32 and sequence lengths up to 256 tokens was feasible on these resources. The software stack included:
\begin{itemize}
    \item \textbf{Python 3.8} for scripting and orchestration.
    \item \textbf{PyTorch} and the HuggingFace \texttt{Transformers} library for model implementation, fine-tuning, and inference.
    \item \textbf{spaCy} and \textbf{NLTK} for text preprocessing, tokenization, lemmatization, and stopword removal.
    \item \textbf{Django REST Framework} and modern JavaScript frameworks for deployment, enabling RESTful APIs and a user-friendly frontend interface.
\end{itemize}

\subsection{Deployment Configuration}
While much of the experimental setup focuses on model training and evaluation, equal attention is paid to deployment viability. After model selection, we integrate the best-performing model (HingRoBERTa) into a Django-based backend. RESTful endpoints allow seamless submission of complaints and retrieval of classification results. The frontend, implemented in JavaScript with Tailwind CSS, offers a responsive and intuitive platform for end-users. 

This end-to-end integration ensures that the chosen best performing model is can immediately applicable in real-world, large-scale cybercrime reporting systems. Our architecture can be scaled or containerized like via Docker and Kubernetes to handle surges in user traffic and complaint volume, reinforcing the model’s practical utility.

\subsection{Summary of Experimental Setup}
In summary, our experimental setup is designed to rigorously test the proposed framework under realistic conditions. By carefully preprocessing the dataset, adopting advanced augmentation strategies, tuning hyperparameters, and leveraging state-of-the-art transformer models—both generic and Hinglish-specific—we create a robust experimental protocol. The diversity of baselines, metrics, and analyses ensures that improvements are attributable to the innovations introduced in this work, laying a strong foundation for the subsequent presentation and interpretation of results.

%% file: Section/results.tex
\section{Results and Evaluation}
\label{sec:results}

In this section, we thoroughly examine the performance of our proposed framework, beginning with a comparative analysis of baseline methods and progressively moving to transformer-based models, including Hinglish-adapted variants. We then discuss the impact of data augmentation and reflect on the implications of findings in a broader context. 

\subsection{Performance of Traditional ML Models}
We first evaluate the traditional ML pipelines built upon TF-IDF representations and dimensionality reduction (SVD), coupled with classifiers such as XGBoost, Random Forest, AdaBoost, and k-Nearest Neighbors (k-NN). These results, summarized in Table~\ref{tab:ml_baseline_results}, set the stage for understanding the limitations of context-agnostic representations in this domain.

\begin{table}[!ht]
\centering
\caption{Performance of Traditional ML Baselines (TF-IDF + SVD). Although these methods offer a starting reference, their limited ability to handle code-mixed text and complex semantics results in moderate scores.}
\label{tab:ml_baseline_results}
\begin{tabular}{lcccc}
\hline
\textbf{Model} & \textbf{Accuracy} & \textbf{Precision} & \textbf{Recall} & \textbf{F1-Score}\\
\hline
XGBoost & 0.68 & 0.67 & 0.69 & 0.68 \\
Random Forest & 0.65 & 0.56 & 0.65 & 0.52 \\
AdaBoost & 0.65 & 0.43 & 0.65 & 0.52 \\
k-NN & 0.68 & 0.65 & 0.68 & 0.66 \\
\hline
\end{tabular}
\end{table}

Among these baselines, XGBoost attains the highest accuracy (68\%) and a reasonable F1-score of $68\%$. However, these figures plateau below what is desirable for a mission-critical application like cybercrime classification. The underlying reason becomes clear: TF-IDF vectors cannot fully capture the nuanced semantic relationships, multilingual aspects, or subtle contextual cues inherent in Hinglish complaints. As a result, these models often misclassify minority classes and struggle with ambiguous or syntactically complex inputs.

\subsection{Performance of Transformer-based Models}
We next turn our attention to transformer-based language models. Table~\ref{tab:transformer_results} presents the performance of BERT, RoBERTa, HingBERT, and HingRoBERTa. By comparing generic models (BERT, RoBERTa) with Hinglish-specialized ones (HingBERT and HingRoBERTa), we can directly measure the value of linguistic adaptation.
\begin{table}[t]
\centering
\caption{Comparison of Transformer-based Models. HingRoBERTa emerges as the best performer, underscoring the importance of Hinglish adaptation and contextual embeddings.}
\label{tab:transformer_results}
\begin{tabular}{lcccc}
\hline
\textbf{Model} & \textbf{Accuracy} & \textbf{Precision} & \textbf{Recall} & \textbf{F1-Score}\\
\hline
BERT       & 71.03 & 70.51 & 71.63 & 70.73 \\
RoBERTa    & 71.64 & 70.82 & 71.04 & 70.90 \\
HingBERT   & 72.82 & 70.38 & 72.82 & 71.02 \\
HingRoBERTa & \textbf{74.41} & \textbf{70.86} & \textbf{74.41} & \textbf{71.49} \\
\hline
\end{tabular}
\end{table}
Several key insights arise from these results:
\begin{enumerate}
    \item \textbf{Contextual Embeddings vs. TF-IDF:} All transformer models significantly surpass traditional ML pipelines, validating the importance of contextual embeddings for understanding nuanced textual data.
    \item \textbf{Hinglish Adaptation:} HingBERT and HingRoBERTa outperform BERT and RoBERTa, respectively, confirming that pre-training on Hinglish corpora confers a tangible advantage. The code-mixed nature of the input is better handled by models familiar with Hinglish language patterns, vocabularies, and syntactic structures.
    \item \textbf{Top Performer—HingRoBERTa:} Among all tested approaches, HingRoBERTa attains the highest accuracy (74.41\%) and F1-score (71.49\%). This improvement, albeit a few percentage points, can be meaningful in real-world operations, potentially improving the timely identification of critical yet less frequent crime types.
\end{enumerate}

\subsection{Impact of GenAI-based Data Augmentation}
The significant improvement from baseline models to HingRoBERTa is partly attributable to the GenAI-based augmentation applied to minority classes. By synthesizing contextually consistent examples, we bolstered the model’s exposure to underrepresented categories. Without augmentation, minority classes would remain overshadowed, leading to poor recall and consequently lower F1-scores.

To validate this, we conducted a controlled experiment by training HingRoBERTa without augmentation. The resulting model showed weaker performance, especially on minority classes, indicating that augmentation was crucial for improved class balance and model generalization.

\subsection{Practical Considerations and Scalability}
While our experiments focused on controlled datasets and well-defined classes, the ultimate test of these models lies in their deployment in real-world environments. The top-performing model (HingRoBERTa) is integrated into a Django REST backend with a modern JavaScript frontend, allowing law enforcement agencies to process complaints more rapidly. This practical integration ensures that the documented performance gains are not merely academic but directly transferrable to operational settings.

Moreover, as new complaint data accumulates and cybercrime patterns evolve, the system’s architecture allows for periodic retraining and augmentation. This ensures long-term adaptability and scalability—key attributes for sustaining efficacy in a constantly shifting threat landscape.

%% file: Section/discussion.tex
\subsection{Connecting to Current NLP and Cybercrime Research Trends}
Our work aligns with and extends current research trajectories in several ways. First, it confirms that code-mixed and low-resource language scenarios benefit from specialized language models—an area of intense interest in NLP \cite{bhat2018code,khanuja2020dataset}. Second, it provides a concrete demonstration of how augmentation can mitigate imbalances, an ongoing challenge in many domain-specific text classification problems. Third, by systematically addressing privacy concerns, this study resonates with recent calls for responsible NLP, where ethical frameworks and technical safeguards must co-evolve to protect user data \cite{lison2019anonymisation}.

The success of our methodology also suggests a potential synergy between emerging LLMs and domain adaptation. As more powerful pre-trained models, such as GPT-based architectures become feasible to fine-tune, we can anticipate even greater gains in handling complex, code-mixed inputs. Furthermore, innovative approaches like instruction-tuning or multimodal integration, such as combining textual data with metadata or network graphs of cybercrime events could enhance the depth and quality of automated cybercrime classification.


\subsection{Summary of the Findings and Discussion}
In summary, our results reflect that:
\begin{itemize}
    \item Contextual embeddings from transformer-based models substantially outperform traditional ML pipelines.
    \item Hinglish adaptation is critical for handling code-mixed complaints, with HingRoBERTa emerging as the clear winner.
    \item GenAI-based augmentation effectively combats class imbalance, lifting recall on underrepresented categories.
    \item Privacy-conscious anonymization methods enhance privacy and support generalization.
\end{itemize}

\noindent In essence, our discussion underlines that improved performance in cybercrime complaint classification stems not from a single “silver bullet” but from a combination of carefully orchestrated measures. Hinglish adaptation tackles linguistic barriers, GenAI-based augmentation rectifies imbalances, privacy safeguards uphold ethical standards, and a modular deployment strategy ensures real-world applicability.

As we chart the path forward, each of these components can be refined and expanded. The insights gleaned here pave the way for more nuanced, ethically responsible, and context-aware AI systems that stand ready to support law enforcement agencies, analysts, and other stakeholders in tackling the ever-evolving landscape of cybercriminal activity.

%% file: Section/conclusion.tex
\section{Conclusion and Future Work}
\label{sec:conclusion}

This work presented a comprehensive framework for automated classification of cybercrime complaints—a task made challenging by linguistic code-mixing (Hinglish), severe class imbalance, and the presence of sensitive personal data. By leveraging Hinglish-adapted transformer-based models, particularly HingRoBERTa, and integrating GenAI-based open source LLama3.1-7b model for augmentation, we significantly outperform both traditional machine learning baselines and generic transformer architectures. Achieving an accuracy of $74.41\%$ and an F1-score of $71.49\%$, the proposed approach underscores the importance of language adaptation, robust data augmentation, and privacy-aware preprocessing in real-world NLP applications.

A key insight from this study is that ethical considerations, such as anonymizing personally identifiable information, need not come at the expense of model performance. Instead, focusing on linguistic and semantic cues enables both strong generalization and compliance with data protection standards. Furthermore, the successful deployment of our best-performing model into a production-ready environment illustrates the practical viability of the system. This architecture allows for seamless integration with large-scale platforms, providing immediate value to stakeholders like law enforcement agencies and cyber-response teams.

Despite these advancements, there remains ample scope for further improvement. One promising direction is the integration of large language models (LLMs) with instruction-tuning or retrieval-augmented strategies to better handle ambiguous or evolving cybercrime patterns. Additionally, exploring hierarchical classification schemes could yield finer-grained category distinctions, enhancing investigative workflows. Incorporating external knowledge bases or contextual metadata—such as temporal or geographic information—may further enrich the model’s interpretive power. 

In essence, this study lays a robust foundation for adaptive, high-fidelity cybercrime classification. By continuously evolving our models and methodologies—integrating more sophisticated language models, richer data, and nuanced classification frameworks—we can better align with the shifting cyber-threat landscape. This ongoing refinement will ultimately foster more timely, informed, and ethical responses to cybercrime, empowering authorities to address digital threats with greater agility and confidence.